\documentclass[aps, prd, amsmath, floats, floatfix, twocolumn, nofootinbib, superscriptaddress]{revtex4-1}
\usepackage[pdftex]{graphicx}
\usepackage{color}
\usepackage{latexsym}
\usepackage[colorlinks]{hyperref}

\newcommand{\be}{\begin{eqnarray}}
\newcommand{\ee}{\end{eqnarray}}

\begin{document}

\title{Constraining the Johannsen deformation parameter $\epsilon_3$ with black hole X-ray data}

\author{Ashutosh~Tripathi}
\affiliation{Center for Field Theory and Particle Physics and Department of Physics, Fudan University, 200438 Shanghai, China}

\author{Askar~B.~Abdikamalov}
\affiliation{Center for Field Theory and Particle Physics and Department of Physics, Fudan University, 200438 Shanghai, China}

\author{Dimitry~Ayzenberg}
\affiliation{Center for Field Theory and Particle Physics and Department of Physics, Fudan University, 200438 Shanghai, China}

\author{Cosimo~Bambi}
\email[Corresponding author: ]{bambi@fudan.edu.cn}
\affiliation{Center for Field Theory and Particle Physics and Department of Physics, Fudan University, 200438 Shanghai, China}

\author{Sourabh~Nampalliwar}
\affiliation{Theoretical Astrophysics, Eberhard-Karls Universit\"at T\"ubingen, 72076 T\"ubingen, Germany}

\begin{abstract}
We extend our previous work on tests of the Kerr black hole hypothesis using X-ray reflection spectroscopy and we infer observational constraints on the Johannsen deformation parameter $\epsilon_3$. We analyze a \textsl{NuSTAR} observation of the stellar-mass black hole in GS~1354--645, \textsl{Suzaku} data of the supermassive black holes in Ark~564, 1H0419--577, and PKS~0558--504, and some simultaneous \textsl{XMM-Newton} and \textsl{NuSTAR} observations of the supermassive black hole in MCG--6--30--15. All our measurements of $\epsilon_3$ are consistent with the Kerr metric and we find quite strong constraints from Ark~564 and MCG--6--30--15. We discuss the implications of our results and the next steps to improve our tests.
\end{abstract}

\maketitle


\section{Introduction}

Einstein's theory of general relativity is a fundamental pillar in modern physics. While the theory has been extensively tested for weak gravitational fields~\cite{will}, its strong gravity regime is still largely unexplored. Astrophysical black holes are an ideal laboratory for testing the predictions of Einstein's gravity in the strong field regime~\cite{r1,r2,r3,r4,r5}. There are indeed a large number of gravity theories that have the same predictions as general relativity in the weak field limit while they have black holes different from those expected in general relativity.

The spacetime metric of astrophysical black holes is thought to be well described by the Kerr solution of Einstein's gravity~\cite{b1,b2} and be completely characterized by two parameters, representing, respectively, the mass and the spin angular momentum of the compact object~\cite{h1,h2,h3}. Testing the Kerr nature of astrophysical black holes is thus a test of Einstein's gravity in the strong field regime.

X-ray reflection spectroscopy is potentially a powerful technique for testing the Kerr black hole hypothesis. Within the disk-corona model, an accreting black hole is surrounded by a geometrically thin and optically thick accretion disk~\cite{b1,b3}. Every point of the disk has a thermal blackbody-like spectrum, which becomes a multi-temperature blackbody spectrum when we consider the whole disk. The temperature at the inner edge of the accretion disk mainly depends on the black hole mass and on the mass accretion rate, and it is usually around 0.1-1~keV for stellar mass black holes and around 1-10~eV for supermassive black holes. Thermal photons from the accretion disk can inverse Compton scatter off free electrons in the ``corona'', which is a hotter ($\sim 100$~keV), often compact and optically thin, cloud in the vicinity of the black hole but whose exact morphology is currently unknown. Comptonized photons from the corona have a power law spectrum with an exponential high energy cutoff. A fraction of these photons can illuminate the disk, producing a reflection component~\cite{ref1,ref2,ref3}.

In the rest-frame of the gas of the disk, the strongest features of such a reflection component are usually the iron K$\alpha$ complex, which is a bunch of narrow emission lines at 6.4-6.97~keV depending on the ionization of iron atoms, and the Compton hump at 10-30~keV. Because of the relativistic motion of the gas in the disk and of the propagation of the photons in the strong gravity region around the black hole, the reflection spectrum as observed far from the source is significantly altered, and the iron K$\alpha$ complex usually becomes a very broad feature. In the presence of high quality data and the correct astrophysical model, the study of the disk's reflection spectrum can provide information about the spacetime metric around the compact object~\cite{nk1,nk2,nk3,nk4,nk5,nk6,nk7,nk8}.

Recently, we have constructed the relativistic reflection model {\sc relxill\_nk} to test the Kerr nature of astrophysical black holes~\cite{noi0,noi0b}. {\sc relxill\_nk} is an extension of {\sc relxill}~\cite{relxill1,relxill2,relxill3} to non-Kerr spacetimes. Following the bottom-up approach, {\sc relxill\_nk} employs a parametric black hole metric in which, in addition to the black hole mass and spin angular momentum, the spacetime is described by a number of ``deformation parameters'' introduced to quantify possible deviations from the Kerr geometry. When all deformation parameters vanish, we recover the Kerr metric. From the comparison of the theoretical predictions of {\sc relxill\_nk} with the available X-ray data of a certain astrophysical black hole, we can constrain the value of these deformation parameters and check whether the spacetime metric around the source is consistent with the Kerr solution.

In Refs.~\cite{noi1,noi2,noi3,noi4,noi5,noi6,noi7,noi8}, we have used {\sc relxill\_nk} to analyze X-ray data of a number of stellar-mass and supermassive black holes to constrain the Johannsen deformation parameters $\alpha_{13}$ and $\alpha_{22}$~\cite{tj}: in the Johannsen metric, $\alpha_{13}$ and $\alpha_{22}$ are the two leading order deformation parameters with the strongest impact on the reflection spectrum of the disk. All our measurements are consistent with the Kerr black hole hypothesis and some constraints are surprisingly stringent.

In the present paper, we want to extend our previous work and constrain the Johannsen deformation parameter $\epsilon_3$. We choose some suitable sources from our previous studies. We analyze a \textsl{NuSTAR} observation of the stellar-mass black hole in GS~1354--645, \textsl{Suzaku} observations of the supermassive black holes in Ark~564, 1H0419--577, and PKS~0558--504, and simultaneous \textsl{XMM-Newton} and \textsl{NuSTAR} data of the supermassive black hole in MCG--6--30--15. All our measurements are consistent with a vanishing value of the deformation parameter $\epsilon_3$. Such a result was not obvious {\it a priori} because every deformation parameter has its own impact on the observed reflection spectrum and a different correlation with the other model parameters.

The present manuscript is organized as follows. In Section~\ref{s-metric}, we briefly review the Johannsen metric. In Section~\ref{s-ana}, we present our sources, the spectral analysis, and the constraints on the deformation parameter $\epsilon_3$. Section~\ref{s-dis} is devoted to the discussion of our results and the concluding remarks. Throughout the paper, we adopt the convention $G_{\rm N} = c = 1$ and a metric with signature $(-+++)$.


\section{Johannsen metric \label{s-metric}}

The Johannsen metric has a few remarkable properties that make it suitable/intersting for tests of the Kerr metric with electromagnetic techniques~\cite{tj}: $i)$ it is regular on and outside of the event horizon, $ii)$ it possesses three independent constants of motion (like the Kerr metric), $iii)$ it can somehow approximate some known black hole solutions of modified theories of gravity for suitable choices of the values of its deformation parameters. In Boyer-Lindquist-like coordinates, the line element of the Johannsen metric reads~\cite{tj}
\be\label{eq-jm}
ds^2 &=&-\frac{\tilde{\Sigma}\left(\Delta-a^2A_2^2\sin^2\theta\right)}{B^2}dt^2
+\frac{\tilde{\Sigma}}{\Delta A_5}dr^2+\tilde{\Sigma} d\theta^2 \nonumber\\
&&-\frac{2a\left[\left(r^2+a^2\right)A_1A_2-\Delta\right]\tilde{\Sigma}\sin^2\theta}{B^2}dtd\phi \nonumber\\
&&+\frac{\left[\left(r^2+a^2\right)^2A_1^2-a^2\Delta\sin^2\theta\right]\tilde{\Sigma}\sin^2\theta}{B^2}d\phi^2 \, ,
\ee
where $M$ is the black hole mass, $a = J/M$, $J$ is the black hole spin angular momentum, $\tilde{\Sigma} = \Sigma + f$, and
\be
&&  \Sigma = r^2 + a^2 \cos^2\theta \, , \qquad
\Delta = r^2 - 2 M r + a^2 \, , \nonumber\\
&& B = \left(r^2+a^2\right)A_1-a^2A_2\sin^2\theta \, .
\ee
The functions $f$, $A_1$, $A_2$, and $A_5$ are defined as
\be
f &=& \sum^\infty_{n=2} \epsilon_n \frac{M^n}{r^{n-2}} \, , \\
A_1 &=& 1 + \sum^\infty_{n=0} \alpha_{1n} \left(\frac{M}{r}\right)^n \, , \\
A_2 &=& 1 + \sum^\infty_{n=0} \alpha_{2n}\left(\frac{M}{r}\right)^n \, , \\
A_5 &=& 1 + \sum^\infty_{n=0} \alpha_{5n}\left(\frac{M}{r}\right)^n \, .
\ee
$\{ \epsilon_n \}$, $\{ \alpha_{1n} \}$, $\{ \alpha_{2n} \}$, and $\{ \alpha_{5n} \}$ are four infinite sets of deformation parameters and the Kerr metric is recovered when all deformation parameters vanish. However, in order to recover the correct Newtonian limit and pass Solar System experiments without fine-tuning, some low-order deformation parameters are requested to vanish and the functions $f$, $A_1$, $A_2$, and $A_5$ become~\cite{tj}
\be
f &=& \sum^\infty_{n=3} \epsilon_n \frac{M^n}{r^{n-2}} \, , \\
A_1 &=& 1 + \sum^\infty_{n=3} \alpha_{1n} \left(\frac{M}{r}\right)^n \, , \\
A_2 &=& 1 + \sum^\infty_{n=2} \alpha_{2n} \left(\frac{M}{r}\right)^n \, , \\
A_5 &=& 1 + \sum^\infty_{n=2} \alpha_{5n} \left(\frac{M}{r}\right)^n \, .
\ee
The leading order deformation parameters are thus $\epsilon_3$, $\alpha_{13}$, $\alpha_{22}$, and $\alpha_{52}$.

The deformation parameters $\alpha_{13}$ and $\alpha_{22}$ are the two with the strongest impact on the reflection spectrum~\cite{noi0} and the corresponding observational constraints have been obtained in our previous studies~\cite{noi1,noi2,noi3,noi4,noi5,noi6,noi7,noi8}. The deformation parameter $\alpha_{52}$ only appears in the metric coefficient $g_{rr}$, so it has no impact in the structure of an infinitesimally thin equatorial disk and only affects the photon propagation: eventually it is extremely hard to constrain the value of this parameter using X-ray reflection spectroscopy and presumably we should employ other techniques more sensitive to $g_{rr}$. The deformation parameter $\epsilon_3$ has a moderate effect on the reflection spectrum of accretion disks~\cite{noi0} and has never been constrained as of now: this will be the aim of the present paper. Note that $\epsilon_3$ has a peculiar property with respect to the other leading order deformation parameters: it only affects the motion of massive particles, while it has no effect at all on massless particles like the photons emitted by the disk (this can be easily seen by noticing that $\tilde{\Sigma}$ appears as a conformal factor in front of the metric).

In the analysis presented in this paper, we will ignore spacetimes with pathological properties (spacetime singularities, regions with closed time-like curves, etc.). This requires some restrictions on the values of the spin parameter $a_* = a/M = J/M^2$ and of the deformation parameter $\epsilon_{3}$. For $a_*$, we require
\be
- 1 < a_* < 1 \, .
\ee 
As in the Kerr metric, this is simply the condition for the existence of the event horizon: for $| a_* | > 1$, there is no horizon and the central singularity is naked. For $\epsilon_{3}$, we need~\cite{tj}
\be
\label{eq-constraints}
\epsilon_3 > - \left( 1 + \sqrt{1 - a^2_*} \right)^3 \, .
\ee

\begin{figure}[t]
\begin{center}
\includegraphics[width=8.0cm]{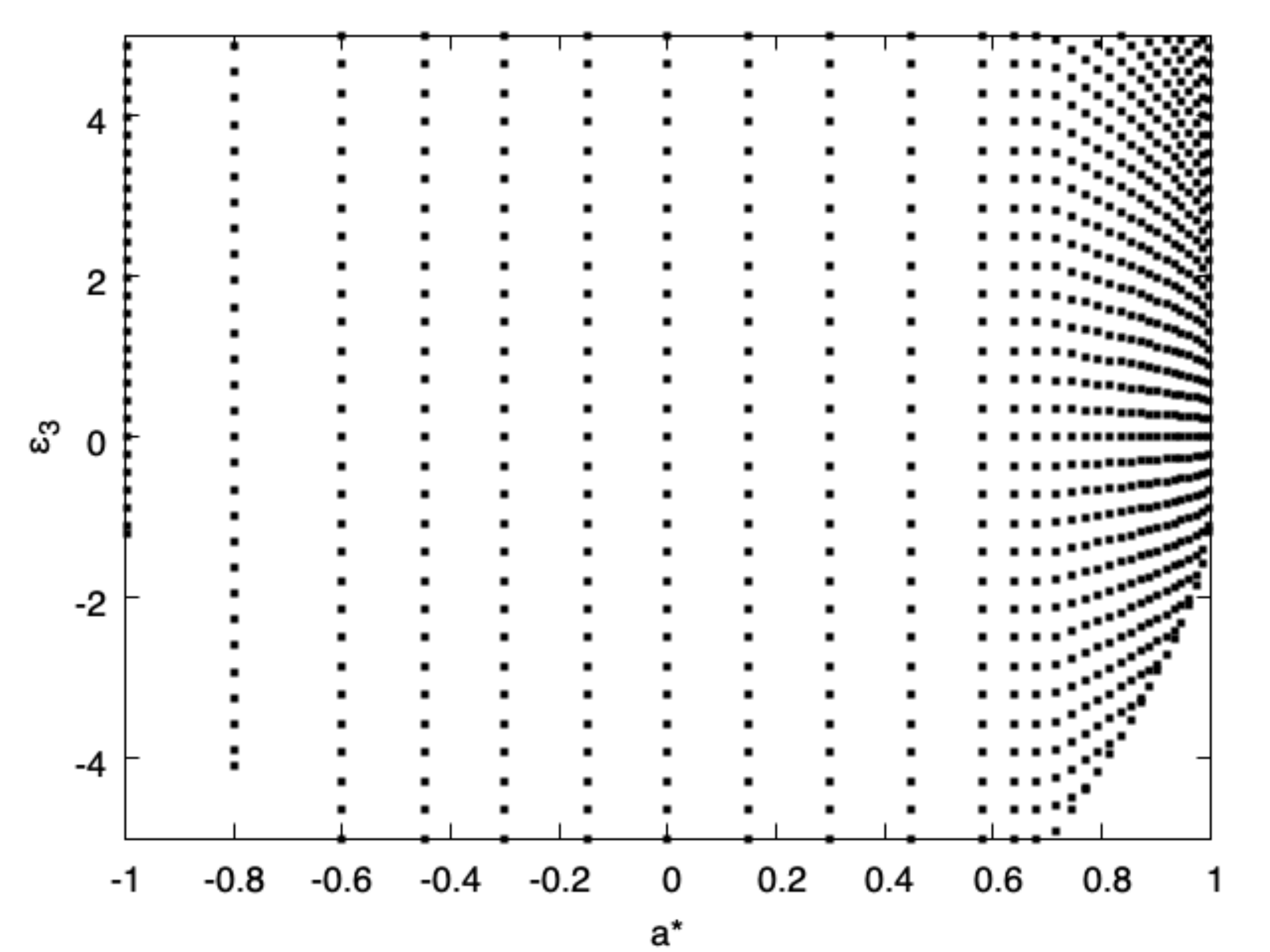}
\end{center}
\vspace{-0.4cm}
\caption{Grid points in the FITS file for the spin parameter $a_*$ and the deformation parameter $\epsilon_3$. \label{f-grid}}
\vspace{0.2cm}
\begin{center}
\includegraphics[width=8.0cm,trim={1.5cm 7.0cm 3.0cm 9.0cm},clip]{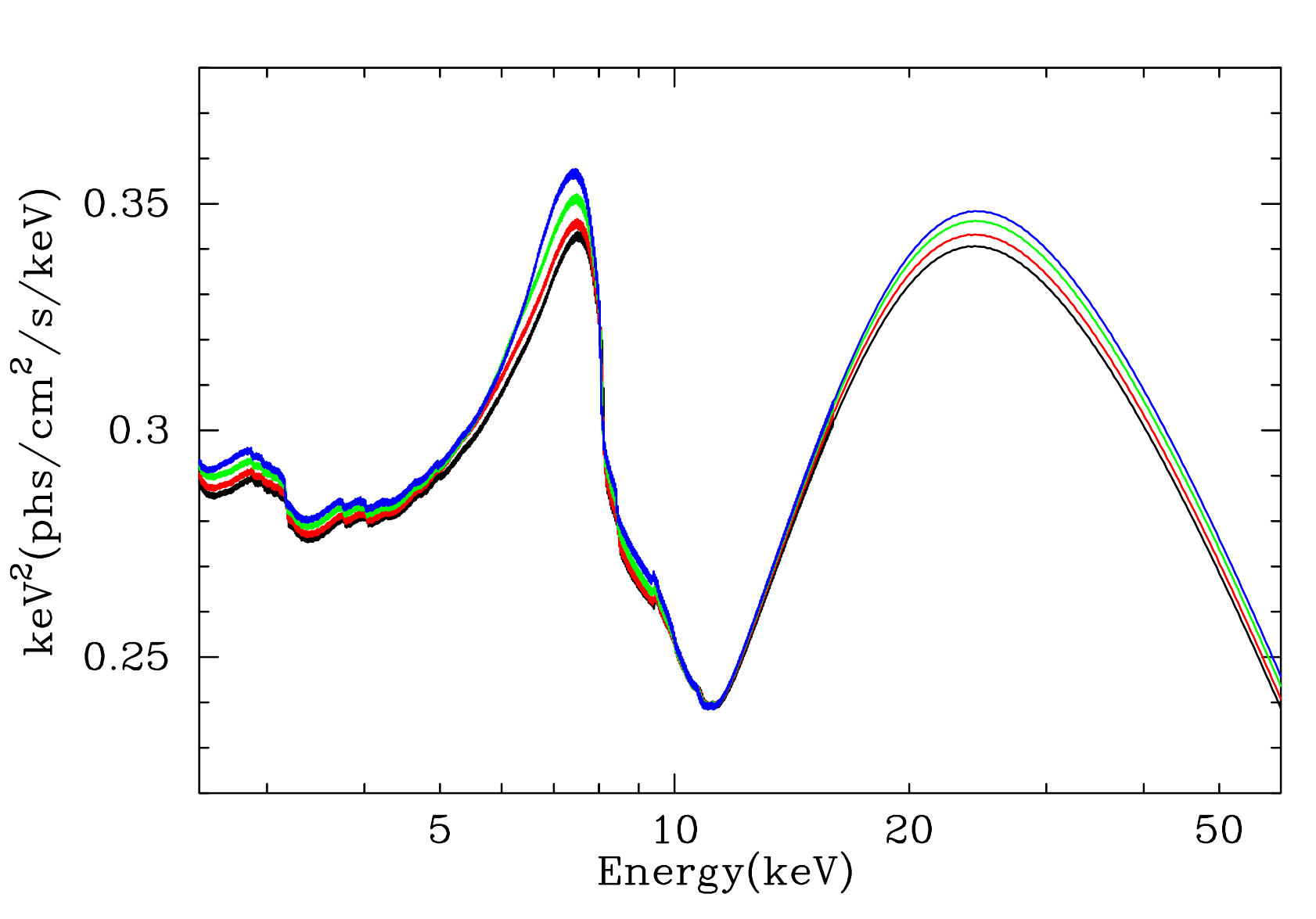}
\end{center}
\vspace{-0.5cm}
\caption{Reflection spectra as calculated by {\sc relxill\_nk} for different values of the deformation parameters $\epsilon_3$: $\epsilon_3 = -1$ (black curve), 0 (red curve), 1 (green curve), and 2 (blue curve). The values of the other model parameters are: $q_{\rm in} = q_{\rm out} = 3$, $a_* = 0.97$, $i = 60^\circ$, $A_{\rm Fe} = 5$, $\log\xi = 3.1$, $\Gamma = 2$, $E_{\rm cut} = 300$~keV. \label{f-e3}}
\end{figure}

The construction of the reflection model to constrain the Johannsen deformation parameter $\epsilon_3$ is equivalent to that for the models with $\alpha_{13}$ and $\alpha_{22}$ and described in Refs.~\cite{noi0,noi0b}. {\sc relxill\_nk} is unchanged and the information related to the spacetime metric are stored in a transfer function, which is tabulated in a FITS file created by a ray-tracing code. The points of the grid on the plane spin vs deformation parameter $\epsilon_3$ are shown in Fig.~\ref{f-grid}. The impact of the Johannsen deformation parameter $\epsilon_3$ on the reflection spectrum is illustrated in Fig.~\ref{f-e3}.

\section{Observational constraints \label{s-ana}}

The aim of this work is to get, for the first time, observational constraints on the deformation parameter $\epsilon_3$. So we will consider sources and observations from our previous tests of the Kerr metric and we will re-analyze those that have shown to be the most suitable/promising to constrain possible deviations from the Kerr geometry. For this reason, this paper will focus on the final constraints on $\epsilon_3$ and the implications of our results, while more details on the data reduction and analysis can be found in our previous work~\cite{noi2,noi4,noi6,noi7}.

\subsection{Sources and observations}

As an example of stellar-mass black hole, we consider GS~1354--645, which was observed by \textsl{NuSTAR} during the outburst of 2015. We consider the observations analyzed in~\cite{noi4} (Obs. ID 90101006004), where we obtained very strong constraints on the deformation parameters $\alpha_{13}$ and $\alpha_{22}$. In~\cite{noi4}, we found that the inner edge of the disk is very close to the compact object and that the disk's inclination angle is quite high, two ingredients that maximize the relativistic effects and are thus useful to get stronger constraints on possible non-Kerr features. \textsl{NuSTAR} data are also particularly suitable for our tests of the Kerr metric, because there is no pile-up problem for bright sources like GS~1354--645 and we have data up to 80~keV.

We also consider three ``bare'' active galactic nuclei from the sources studied in~\cite{noi2,noi7}. These sources have no complicating intrinsic absorption (hence the name ``bare'') and a very simple spectrum, two ingredients that help to reduce the systematic uncertainties and can limit degeneracy among model parameters. We select those sources that have provided strong and well understood constraints in our previous work: Ark~564, 1H0419--577, and PKS~0558--504.

Last, we want to re-analyze the simultaneous observations of \textsl{XMM-Newton} and \textsl{NuSTAR} of 2013 of the supermassive black hole in MCG--6--30--15~\cite{noi6}. MCG--6--30--15 is a very bright source and has a strong reflection spectrum with a very broad iron line. \textsl{XMM-Newton} data permit to have a good energy resolution at the iron line and \textsl{NuSTAR} data cover a wide energy band. Since the source is very variable, we need to group the data into four flux states (low, medium, high, and very-high) and tie in the fit the model parameters that are supposed to be constant over different flux states during the observation, while the model parameters that can vary on short timescales are allowed to change value among different flux states.

Tab.~\ref{tab1} shows the list of sources and observations analyzed in this work.

\begin{table*}
\centering
\vspace{0.3cm}
\begin{tabular}{ccccc}
$\qquad$ Source $\qquad$ & $\qquad$ Mission $\qquad$ & $\qquad$ Observation ID $\qquad$ & $\quad$ Year $\quad$ & $\quad$ Exposure (ks) $\quad$ \\ 
\hline
GS~1354--645 & \textsl{NuSTAR} & 90101006004 & 2015 & 29 \\
\hline
Ark~564 & \textsl{Suzaku} & 702117010 & 2007 & 80 \\
\hline
1H0419--577 & \textsl{Suzaku} & 702041010 & 2007 & 179 \\
\hline
PKS~0558--504 & \textsl{Suzaku} & 701011010 & 2007 & 20 \\
&& 701011020 & 2007 & 19 \\
&& 701011030 & 2007 & 21 \\
&& 701011040 & 2007 & 20 \\
&& 701011050 & 2007 & 20 \\
\hline
MCG--6--30--15 & \textsl{NuSTAR} & 60001047002 & 2013 & 23 \\
&& 60001047003 & 2013 & 127 \\
&& 60001047005 & 2013 & 30 \\ 
& \textsl{XMM-Newton} & 0693781201 & 2013 & 134 \\
&& 0693781301 & 2013 & 134 \\
&& 0693781401 & 2013 & 49 \\
\hline
\end{tabular}
\caption{List of sources and observations analyzed in this work.}
\label{tab1}
\end{table*}

\subsection{Data reduction}

The data reduction of these observations has been already described in our previous work: in Ref.~\cite{noi4} for GS~1354--645, in Ref.~\cite{noi2} for Ark~564, in Ref.~\cite{noi7} for 1H0419--577 and PKS~0558--504, and in Ref.~\cite{noi6} for MCG--6--30--15. Here we only note that we do not use \textsl{Suzaku} data (Ark~564, 1H0419--577, and PKS~0558--504) in the energy range 1.7-2.5~keV because of calibration uncertainties\footnote{\tiny https://heasarc.gsfc.nasa.gov/docs/suzaku/analysis/sical.html} and that we also ignore the energy range 1.5-2.5~keV in the \textsl{XMM-Newton}/EPIC-Pn data (MCG--6--30--15) because of an effect of the golden edge in the response file due to mis-calibration in the long-term charge transfer inefficiency~\cite{andrea}.

\subsection{Spectral analysis and results}

For the spectral analysis, we use XSPEC v12.9.1~\cite{arnaud} and {\sc relxill\_nk} v~1.3.2~\cite{noi0b}\footnote{\tiny http://www.physics.fudan.edu.cn/tps/people/bambi/Site/RELXILL\_NK.html}.

\subsubsection{GS~1354--645}

For GS~1354--645, we obtain already a good fit with a power law and a relativistic reflection spectrum. The XSPEC model is
\be
\text{\sc tbabs*relxill\_nk} \, . \nonumber
\ee
{\sc tbabs} describes the Galactic absorption~\cite{wilms} and the value of the hydrogen column density is frozen to that calculated using~\cite{nH}\footnote{\tiny http://www.swift.ac.uk/analysis/nhtot/}. {\sc relxill\_nk} describes both the primary component from the corona and the relativistic reflection component from the accretion disk~\cite{noi0,noi0b}. The primary component from the corona has two parameters, the photon index $\Gamma$ and the high energy cutoff $E_{\rm cut}$, and are both left free in the fit. The relativistic reflection component from the disk has nine parameters: inner emissivity index $q_{\rm in}$, outer emissivity index $q_{\rm out}$, breaking radius $R_{\rm br}$, spin parameter $a_*$, deformation parameter $\epsilon_3$, inclination angle of the disk $i$, redshift $z$, ionization parameter $\xi$, and iron abundance $A_{\rm Fe}$. Note that here we assume that the inner edge of the accretion disk is at the radius of the innermost stable circular orbit (ISCO), so it is not a model parameter and is determined by $a_*$ and $\epsilon_3$. The reflection fraction, $R_{\rm ref}$, regulates the relative strength between the primary component from the corona and the relativistic reflection component from the disk.

For the emissivity profile of the disk, here and for the other sources, we try both a power law ($q_{\rm in} = q_{\rm out}$ and $R_{\rm br}$ frozen to some value) and a broken power law ($q_{\rm in}$, $q_{\rm out}$, and $R_{\rm br}$ all free), and we choose the latter if the improvement of the fit is enough to justify two extra parameters. When we choose a broken power law and $q_{\rm out}$ is consistent with 3, we repeat the fit with $q_{\rm out}$ frozen to 3, as we interpret our measurement with the presence of a point-like lamppost corona. For GS~1354--645, as already found in~\cite{noi4}, the fit requires a broken power law with very high $q_{\rm in}$ and very low $q_{\rm out}$, so the geometry of the corona is more complicated than a point-like lamppost scenario. All the other parameters of the relativistic reflection component are left free, with the exception of the redshift $z$ which is set to 0 because GS~1354--645 is in our Galaxy and the effect of its relative motion is negligible.

Tab.~\ref{t-fit} reports the best-fit values of the model parameters. The top left panel in Fig.~\ref{f-r1} shows the model and the ratio between the data and the best-fit model of GS~1354--645. The constraints on the spin parameter $a_*$ and the Johannsen deformation parameters $\epsilon_3$ are shown in Fig.~\ref{f-c1}.

\subsubsection{Ark~564}

The \textsl{Suzaku} data of Ark~564 are fitted with the model
\be
\text{\sc tbabs*relxill\_nk} \, . \nonumber
\ee
Both {\sc tbabs} and {\sc relxill\_nk} have been already described in the previous subsection. The difference here is that the redshift of the source is non-negligible and is frozen to the value of its cosmological redshift. Moreover, here and for the other sources with \textsl{Suzaku} data, $E_{\rm cut}$ is frozen to 300~keV, because the \textsl{Suzaku} data cover up to 10~keV and it is impossible to constrain the value of $E_{\rm cut}$. When we fit the data, we find that a broken power law can better describe the emissivity profile of the accretion disk, and that $q_{\rm out}$ is consistent with 3. We thus repeat the fit with $q_{\rm out}$ frozen to 3.

The best-fit values of the model parameters are reported in Tab.~\ref{t-fit}. The top right panel in Fig.~\ref{f-r1} shows the model and the ratio between the data and the best-fit model. The constraints on the spin parameter $a_*$ and the Johannsen deformation parameters $\epsilon_3$ are shown in Fig.~\ref{f-c1}.

\subsubsection{1H0419--577}

The \textsl{Suzaku} data of 1H0419--577 are fitted with the model
\be
\text{\sc tbabs*(relxill\_nk + xillver)} \, . \nonumber
\ee
{\sc xillver} describes a non-relativistic reflection spectrum~\cite{relxill1,relxill2}, which can be interpreted as the spectrum of some cold material at larger distance from the black hole and illuminated by the corona. The values of the model parameters in {\sc xillver} are tied to those of {\sc relxill\_nk}, with the exception of the ionization parameter $\xi$, which is frozen to 0 because the distant reflector is expected to be far from the black hole and its temperature is low. The quality of these data is not very good, and we do not find any substantial difference by assuming a power law or a broken power law for the emissivity profile of the disk, so we choose the former to minimize the number of free parameters.

The best-fit values of the model parameters, the model and the ratio between the data and the best-fit model, and the constraints on the spin parameter $a_*$ and the Johannsen deformation parameters $\epsilon_3$ are shown, respectively, in Tab.~\ref{t-fit}, Fig.~\ref{f-r1}, and Fig.~\ref{f-c1}.

\subsubsection{PKS~0558--504}

For the \textsl{Suzaku} data of PKS~0558--504 we find a good fit with the model
\be
\text{\sc tbabs*(relxill\_nk + zgauss)} \, . \nonumber 
\ee
{\sc zgauss} describes a narrow emission line around 6.95~keV, which can be naturally interpreted as Fe~XXVI. As in the case of Ark~564, a broken power law for the disk's emissivity profile provides a much better fit and we find $q_{\rm out}$ consistent with 3. We thus repeat the fit with $q_{\rm out}$ frozen to 3. If we replace {\sc zgauss} with {\sc xillver} we get a worse fit, so our final model uses {\sc zgauss}.

\begin{figure*}[t]
\begin{center}
\includegraphics[width=8.5cm,trim={0cm 1.8cm 0cm 1.8cm},clip]{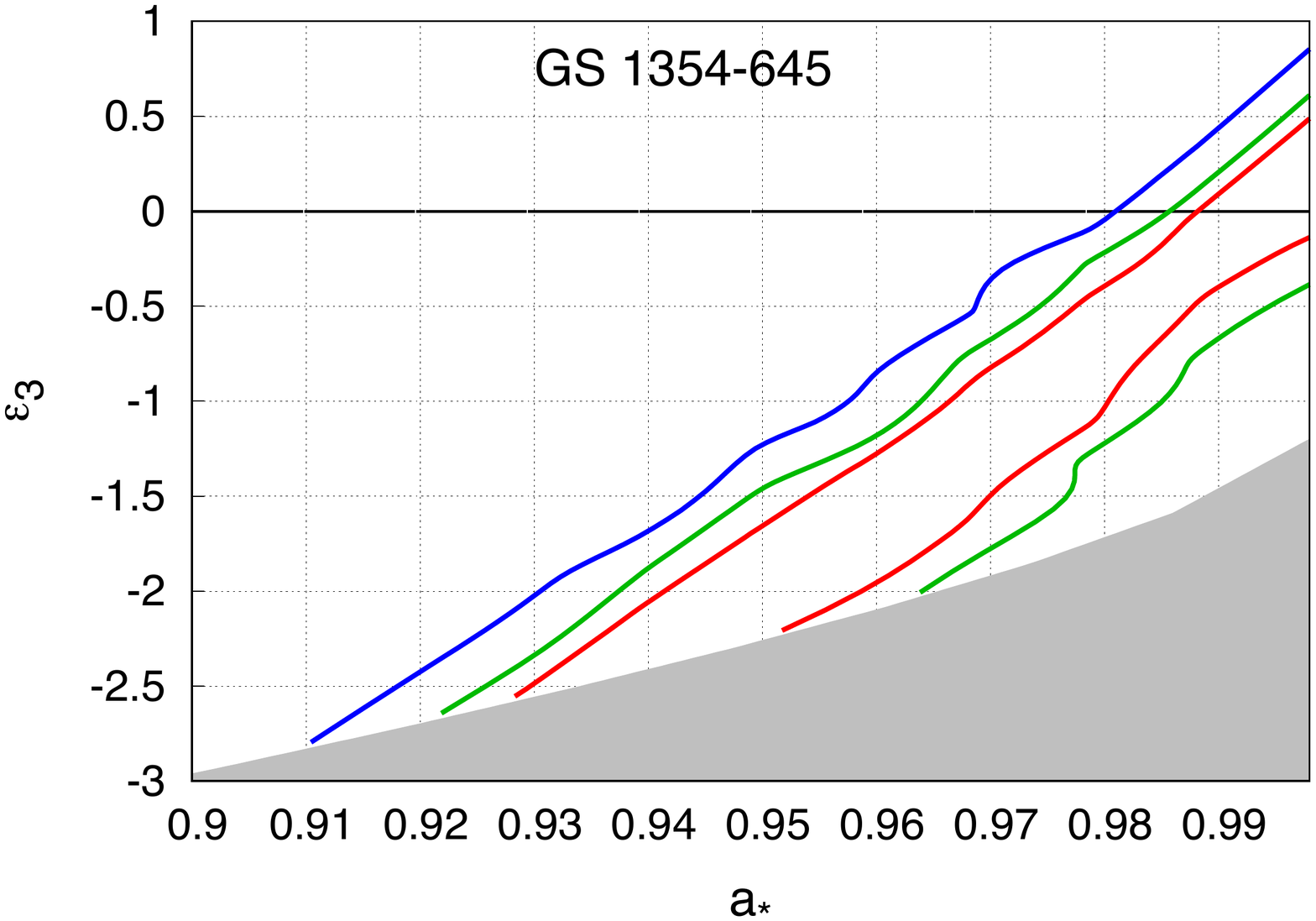}
\includegraphics[width=8.5cm,trim={0cm 1.8cm 0cm 1.8cm},clip]{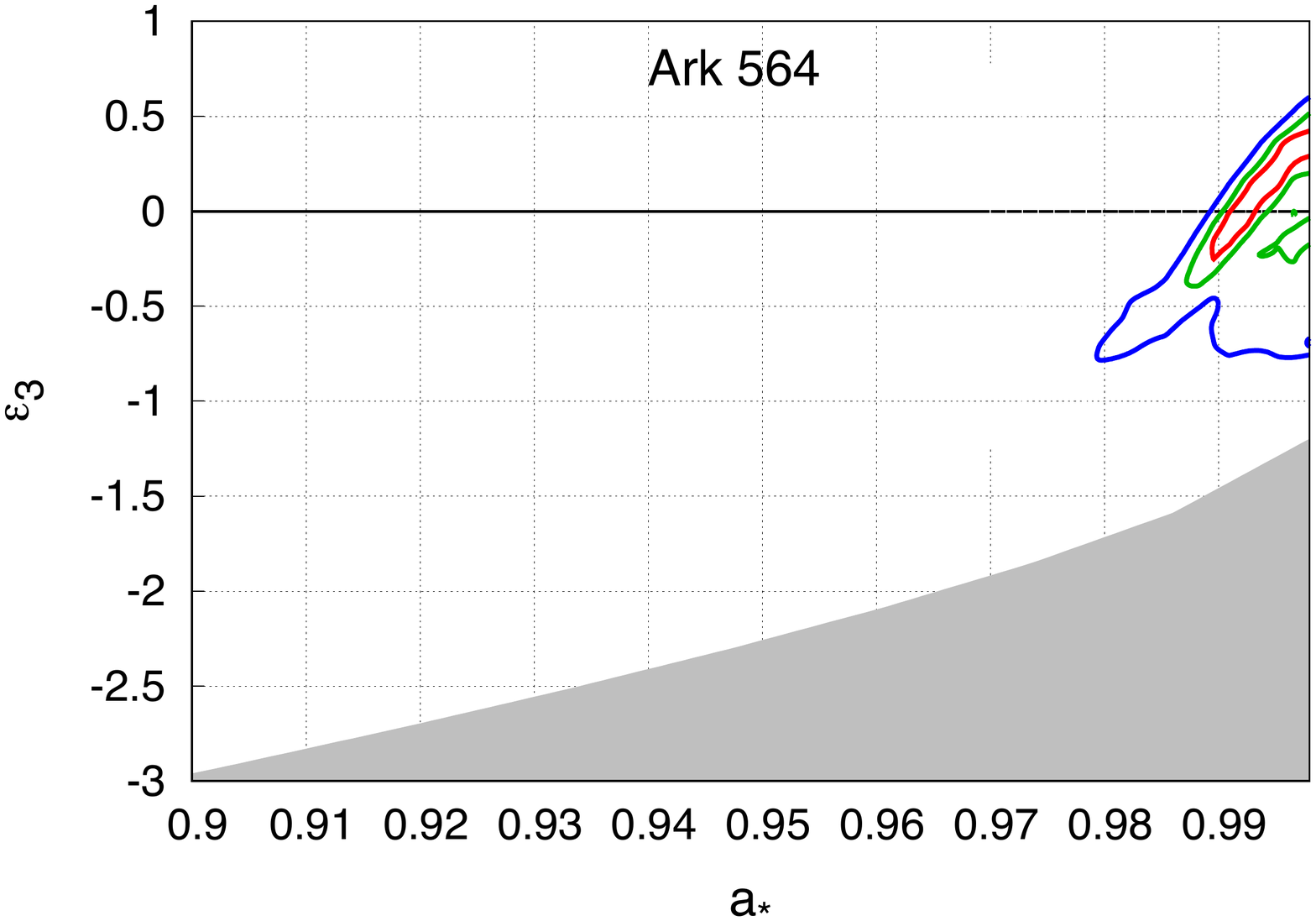}
\includegraphics[width=8.5cm,trim={0cm 1.8cm 0cm 1.8cm},clip]{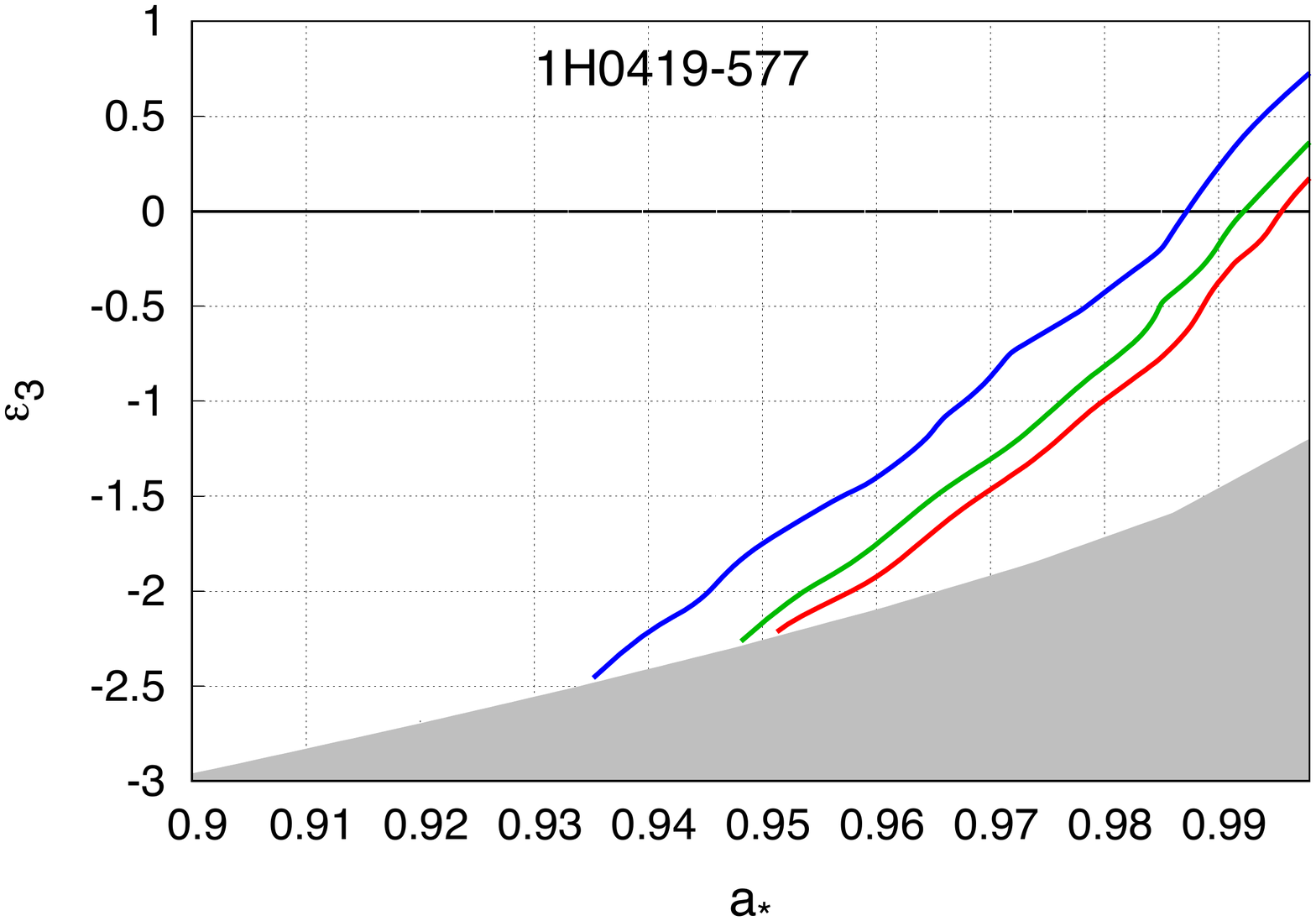}
\includegraphics[width=8.5cm,trim={0cm 1.8cm 0cm 1.8cm},clip]{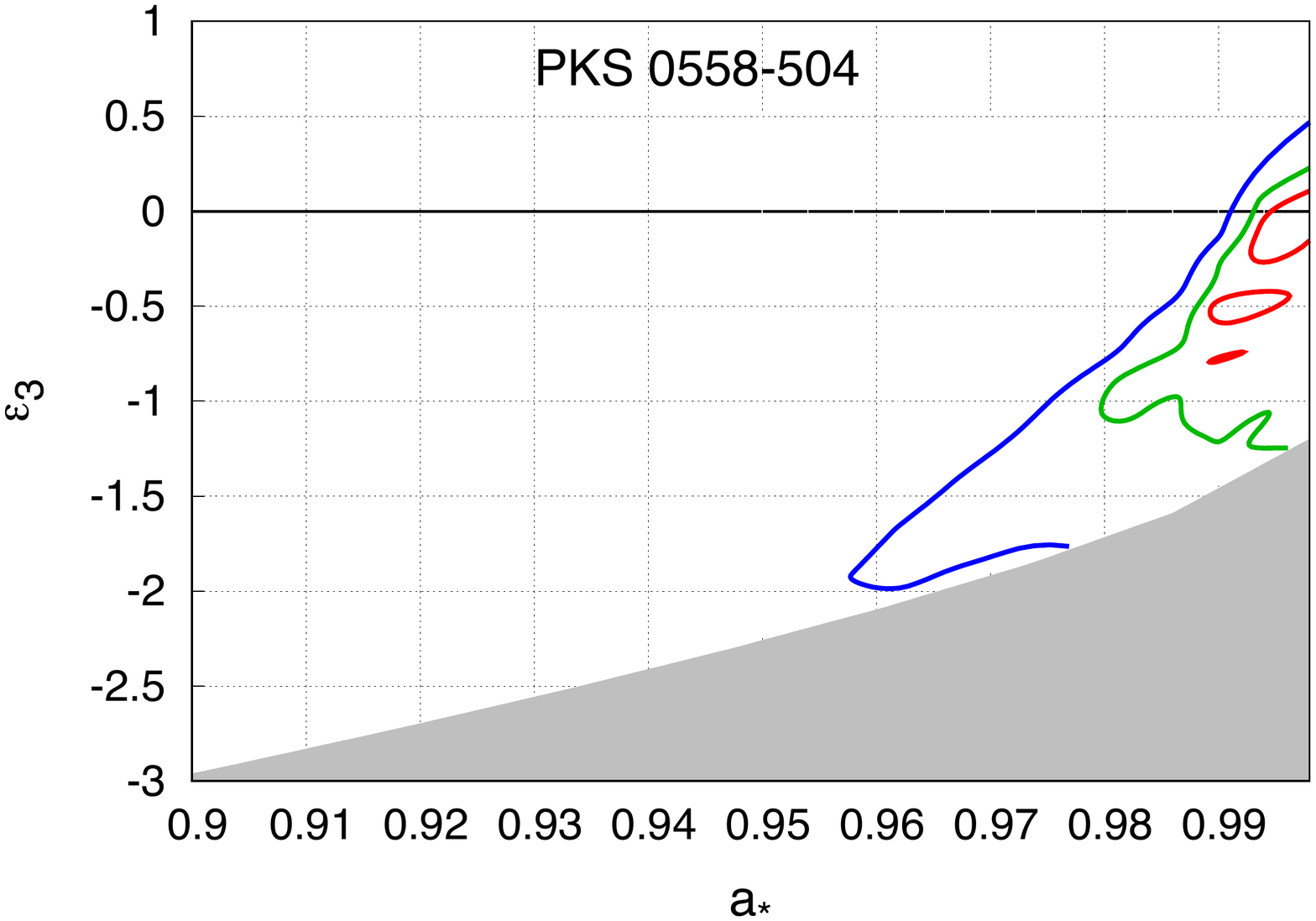}
\end{center}
\vspace{-0.6cm}
\caption{Constraints on the spin parameter $a_*$ and the Johannsen deformation parameters $\epsilon_3$ for the stellar-mass black hole in GS~1354--645 and the supermassive black holes in Ark~564, 1H0419--577, and PKS~0558--504. The red, green, and blue curves are, respectively, the 68\%, 90\%, and 99\% confidence level boundaries for two relevant parameters. The gray region is ignored in our analysis because it violates the constraint in Eq.~(\ref{eq-constraints}).
\label{f-c1}}
\end{figure*}

The results of the analysis of PKS~0558--504 and the corresponding constraints are shown in Tab.~\ref{t-fit}, Fig.~\ref{f-r1}, and Fig.~\ref{f-c1}.

\subsubsection{MCG--6--30--15}

The analysis of MCG--6--30--15 is more complicated than those of the previous sources, because MCG--6--30--15 is quite variable and the spectrum is more complicated due to the presence of some absorbing material crossing the line of sight between the source and us. We proceed as describe in Ref.~\cite{noi6} and we group the data into four sets: low flux state, medium flux state, high flux state, and very-high flux state. We fit the data with the model (see~\cite{noi6} for more details)
\be
&&\text{\sc tbabs*warmabs$_1$*warmabs$_2$*dustyabs} \nonumber\\
&& \qquad \text{\sc *(relxill\_nk + xillver + zgauss + zgauss)} \, . \nonumber 
\ee
{\sc warmabs$_1$} and {\sc warmabs$_2$} describes two ionized absorbers and their tables are generated
with {\sc xstar} v~2.41. {\sc dustyabs} is a neutral absorber and modifies the soft X-ray band due to the presence of dust around the source~\cite{lee99}. Here {\sc zgauss} is used to describe a narrow oxygen emission line around 0.8~keV and a blueshifted narrow oxygen absorption line around 1.2~keV from some relativistic outflow.

\begin{figure*}[t]
\begin{center}
\includegraphics[width=8.5cm,trim={0.5cm 0 3cm 18cm},clip]{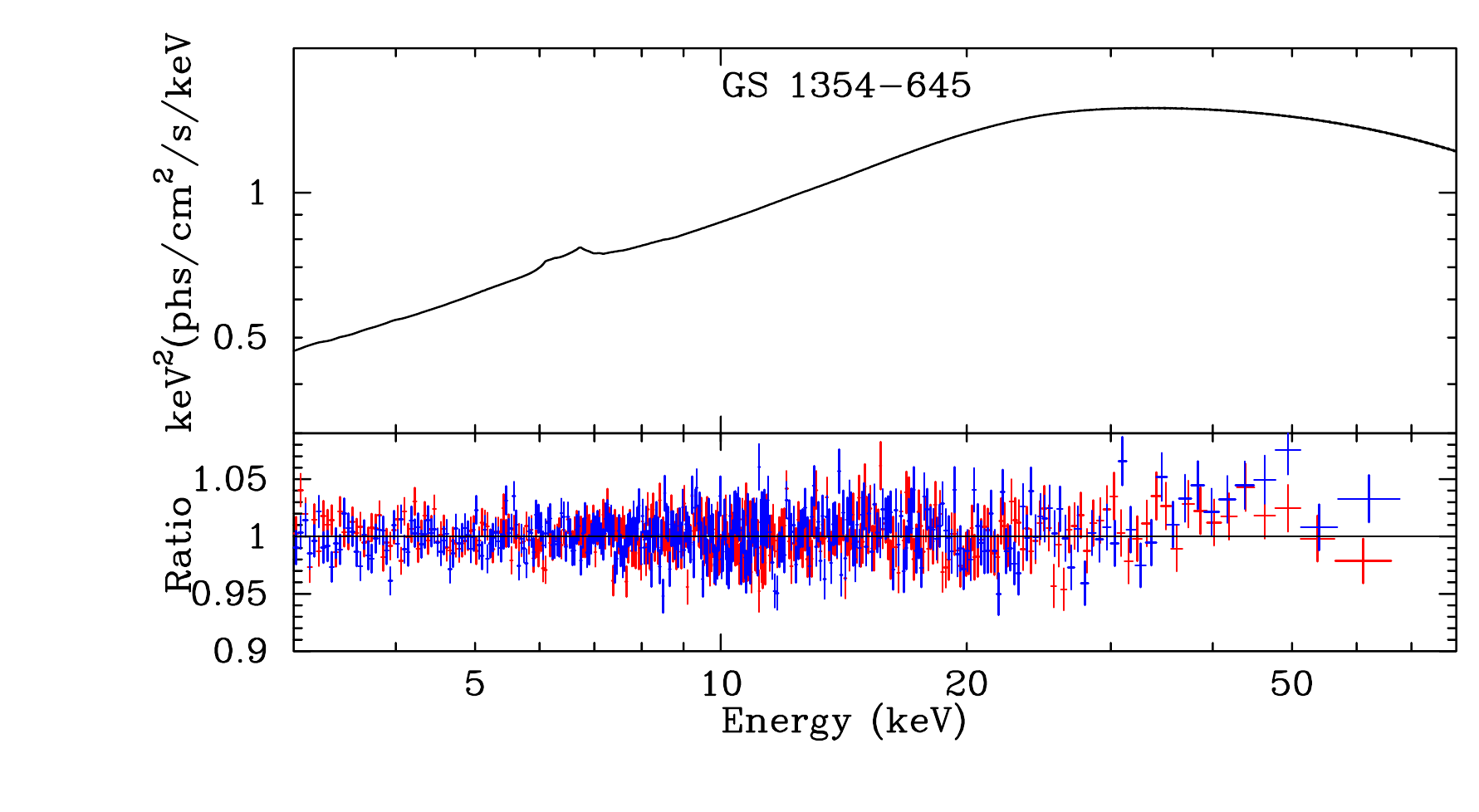}
\includegraphics[width=8.5cm,trim={0.5cm 0 3cm 18cm},clip]{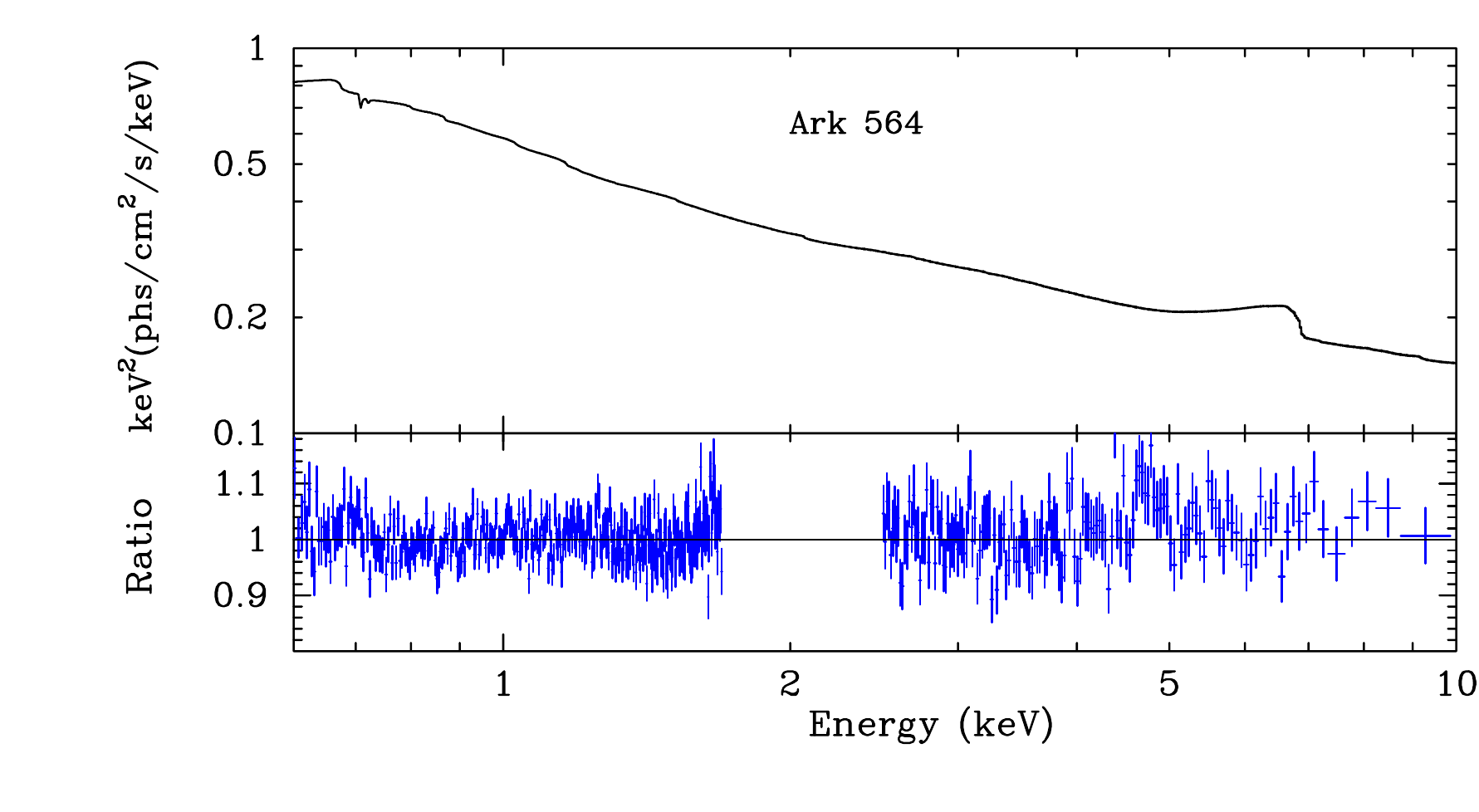} \\
\includegraphics[width=8.5cm,trim={0.5cm 0 3cm 18cm},clip]{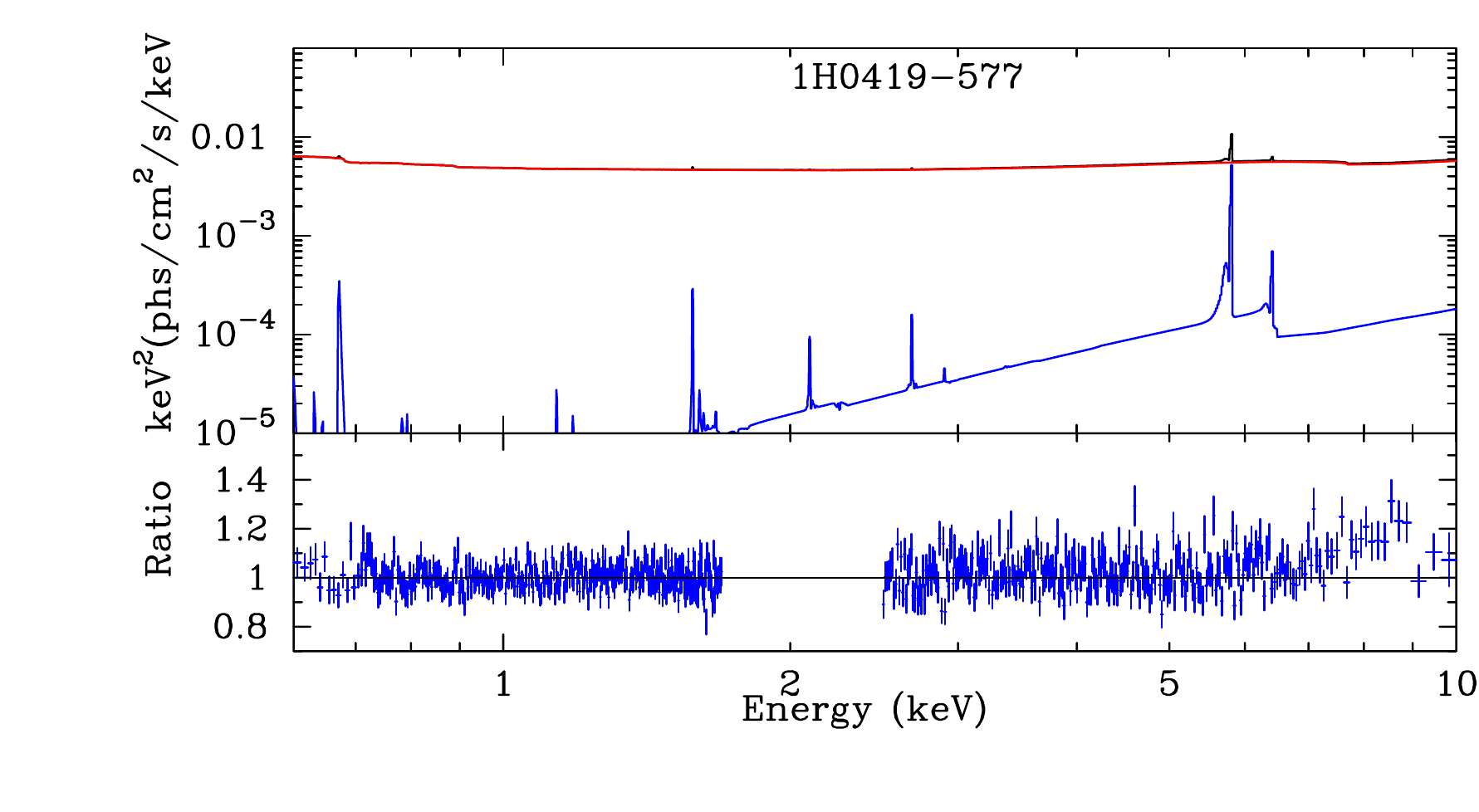}
\includegraphics[width=8.5cm,trim={0.5cm 0 3cm 18cm},clip]{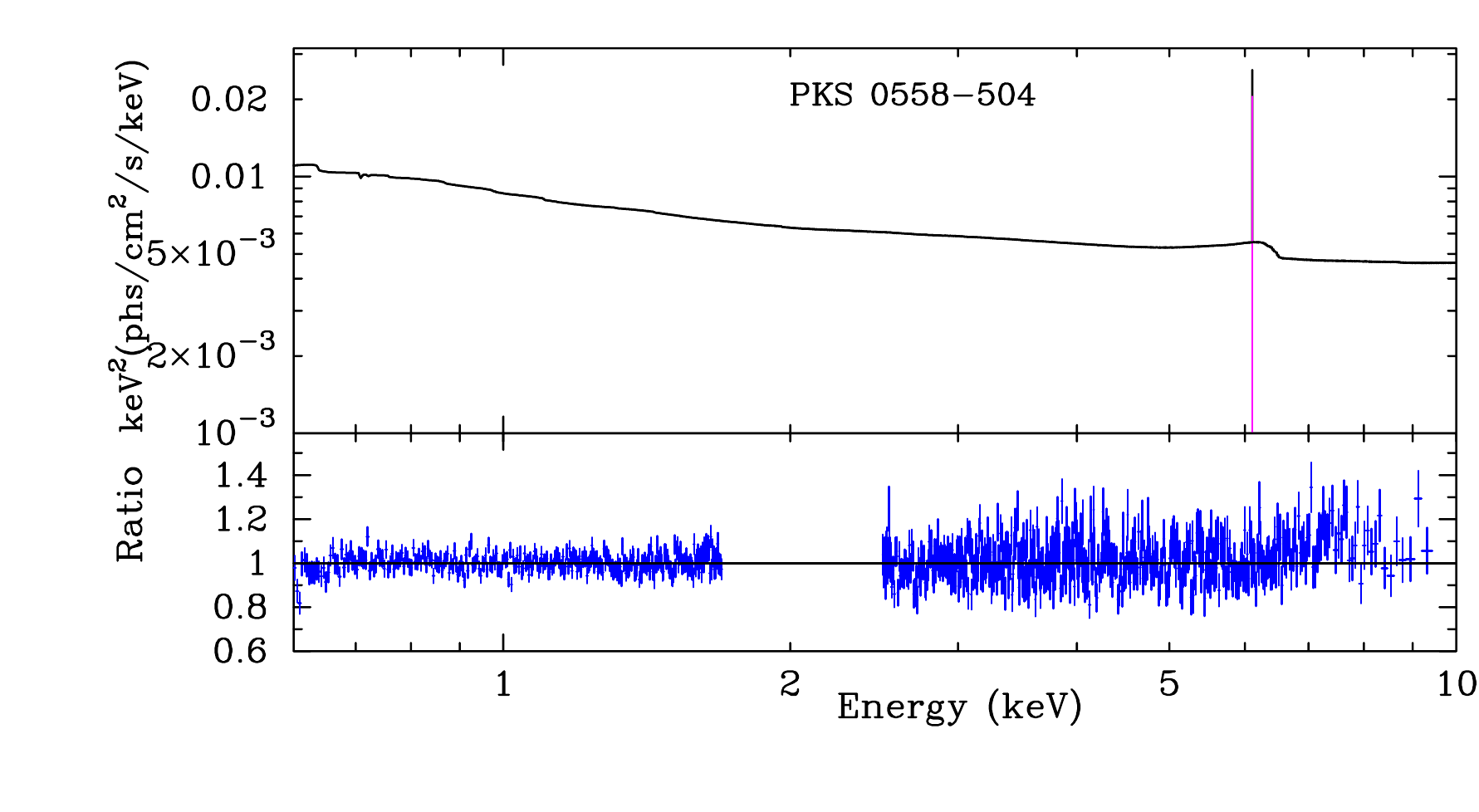}
\end{center}
\vspace{-0.7cm}
\caption{Spectra of the best fit models with the corresponding components (upper panels) and data to best-fit model ratios (lower panels) for the stellar-mass black hole in GS~1354--645 and the supermassive black holes in Ark~564, 1H0419--577, and PKS~0558--504. The total spectrum is in black, the {\sc relxill\_nk} component (power law and relativistic reflection components) is in red, the {\sc xillver} component (non-relativistic reflection component) is in blue, and the narrow emission line is in magenta (for GS~1354--645 and Ark~564, the total model is {\sc relxill\_nk}, so we only see the black curve). In the ratio plot of GS~1354--645, red crosses are for \textsl{NuSTAR}/FPMA and blue crosses are for \textsl{NuSTAR}/FPMB. \label{f-r1}}
\vspace{0.3cm}
\begin{center}
\includegraphics[width=8.5cm,trim={0.5cm 0 3cm 18cm},clip]{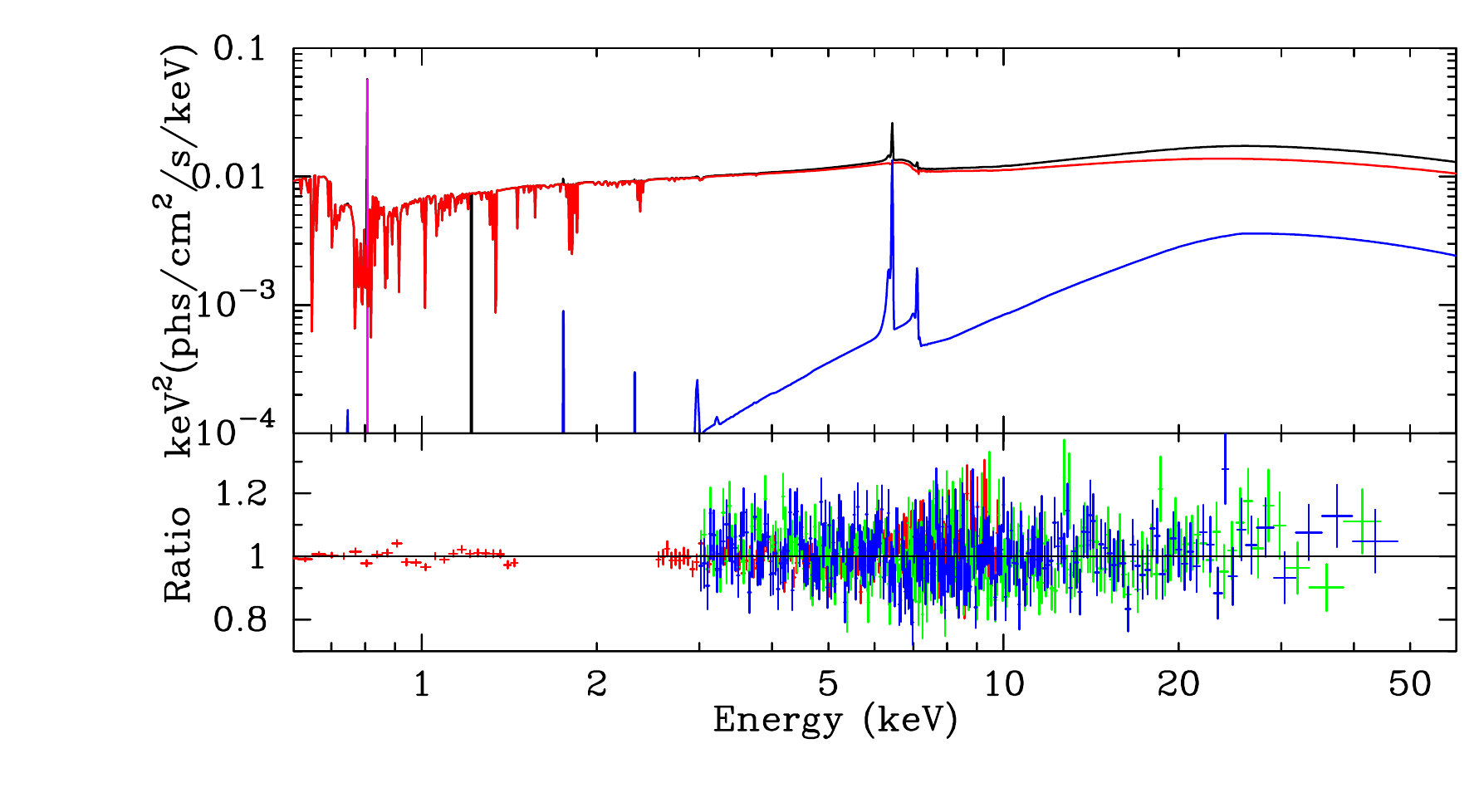}
\includegraphics[width=8.5cm,trim={0.5cm 0 3cm 18cm},clip]{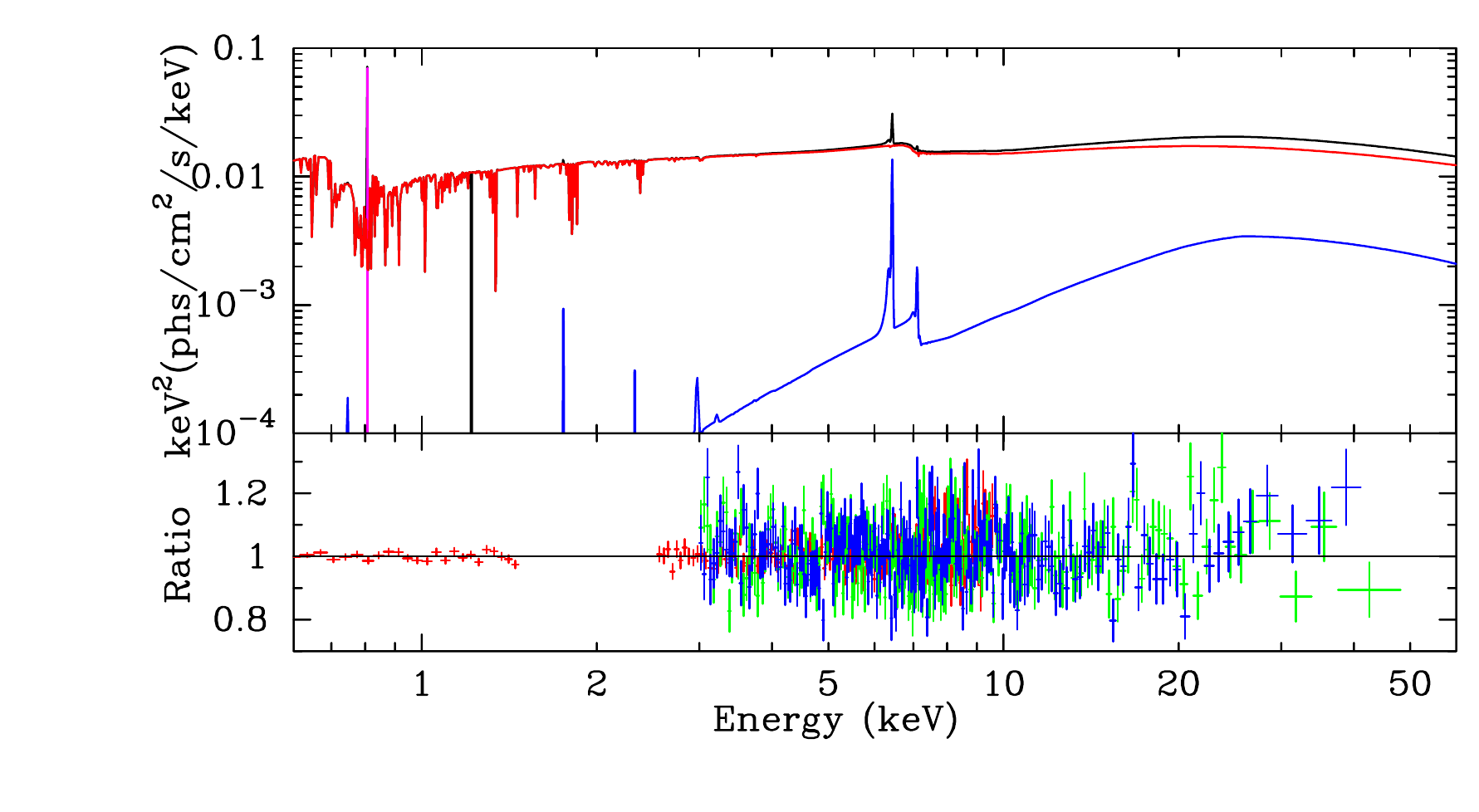} \\
\includegraphics[width=8.5cm,trim={0.5cm 0 3cm 18cm},clip]{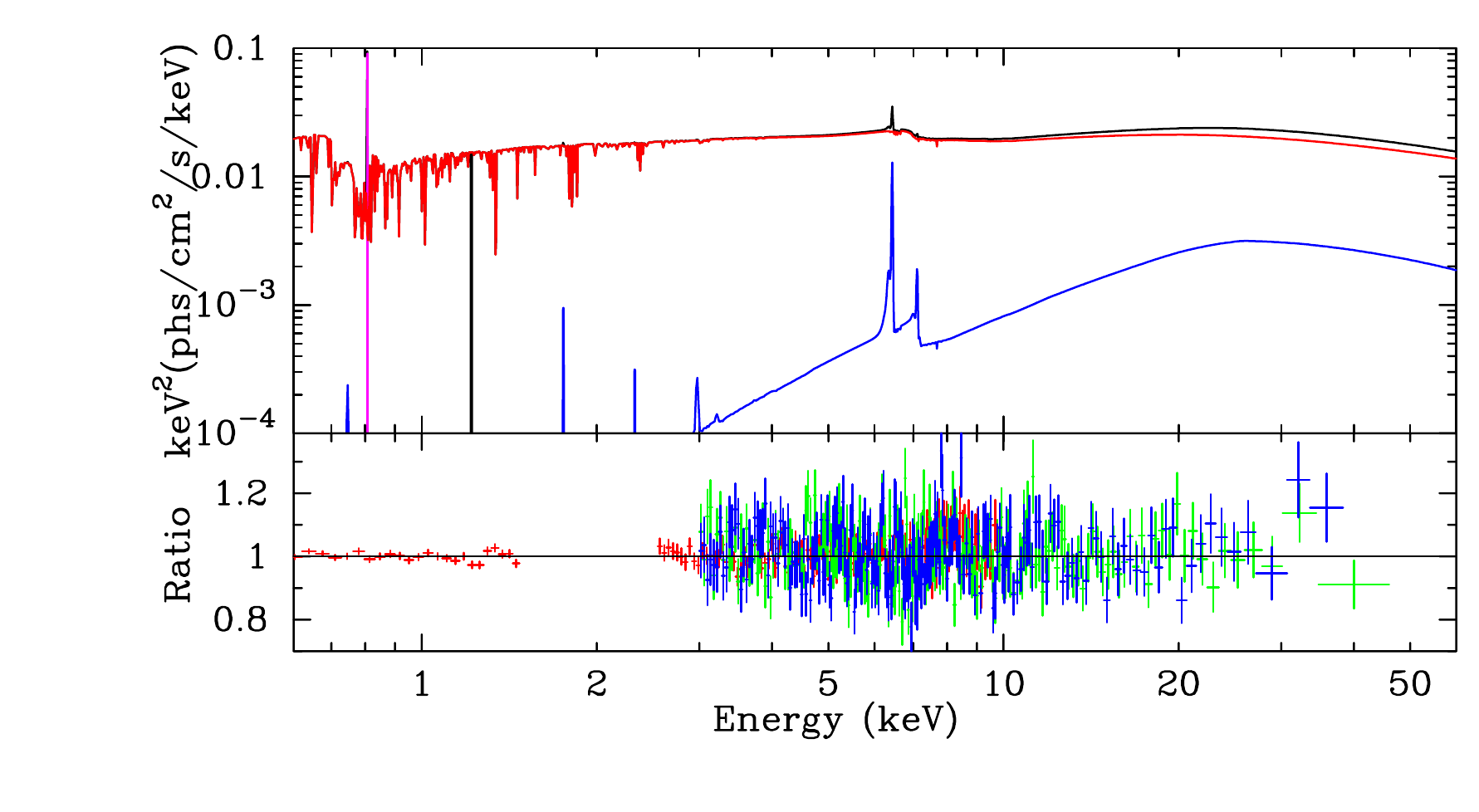}
\includegraphics[width=8.5cm,trim={0.5cm 0 3cm 18cm},clip]{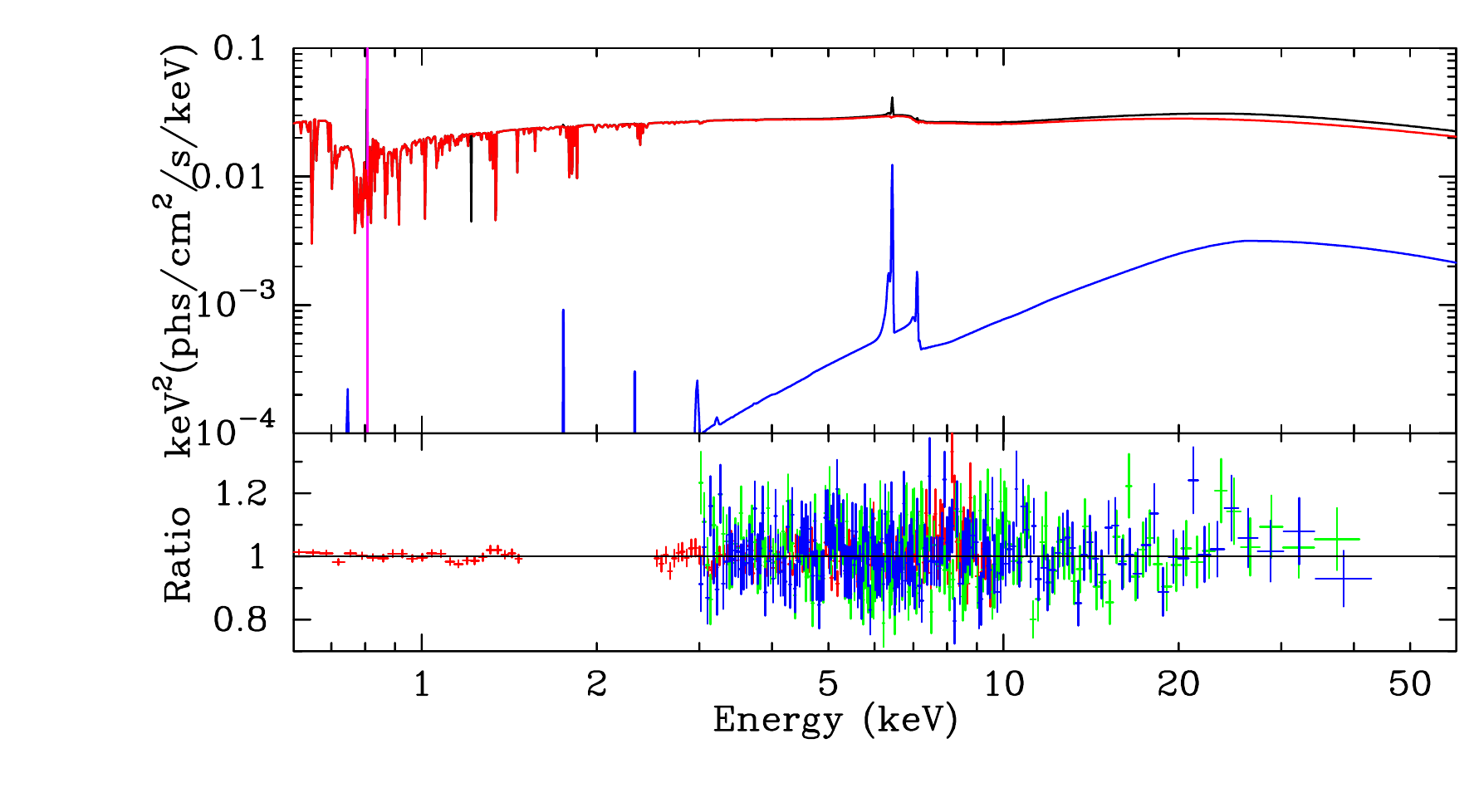}
\end{center}
\vspace{-0.7cm}
\caption{Spectra of the best fit models with the corresponding components (upper panels) and data to best-fit model ratios (lower panels) for the supermassive black holes in MCG--6--30--15 in the low (left top panel), medium (right top panel), high (left bottom panel), and very-high (right bottom panel) flux states. In the model plots, the total spectrum is in black, the {\sc relxill\_nk} component (power law and relativistic reflection components) is in red, the {\sc xillver} component (non-relativistic reflection component) is in blue, and the narrow emission line is in magenta. In the ratio plots, red crosses are for \textsl{XMM-Newton}/EPIC-Pn, green crosses are for \textsl{NuSTAR}/FPMA, and blue crosses are for \textsl{NuSTAR}/FPMB. \label{f-r2}}
\end{figure*}

\begin{figure*}[t]
\begin{center}
\includegraphics[width=8.5cm,trim={0cm 1.8cm 0cm 1.8cm},clip]{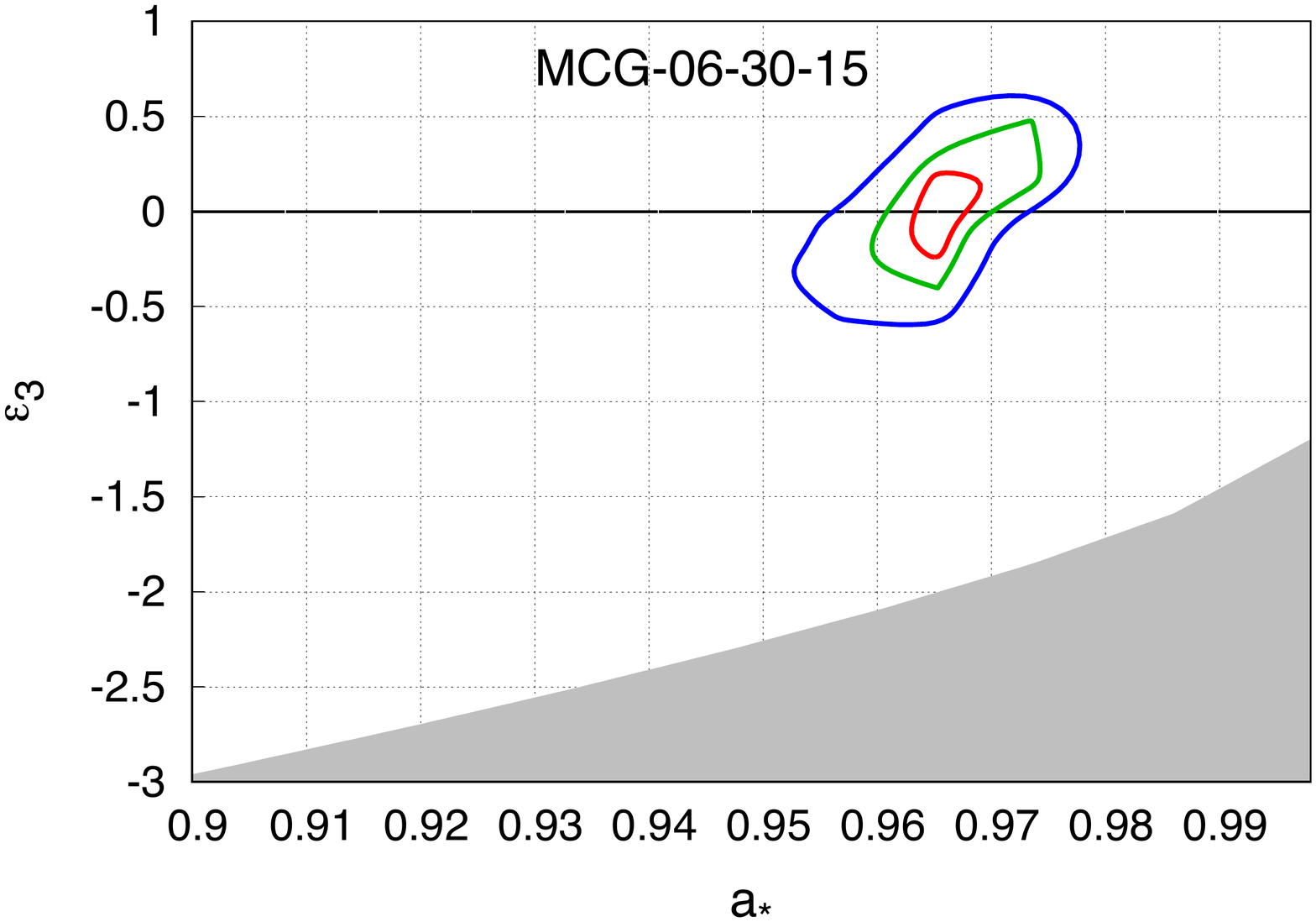}
\end{center}
\vspace{-0.6cm}
\caption{Constraints on the spin parameter $a_*$ and the Johannsen deformation parameters $\epsilon_3$ for the supermassive black hole in MCG--6--30--15. The red, green, and blue curves are, respectively, the 68\%, 90\%, and 99\% confidence level boundaries for two relevant parameters. The gray region is ignored in our analysis because it violates the constraint in Eq.~(\ref{eq-constraints}). \label{f-c2}}
\end{figure*}

The best-fit values of the model parameters are listed in the last column in Tab.~\ref{t-fit}. Note that some model parameters can vary over different flux states and therefore there is a measurement for every flux state (in the order low, medium, high, and very-high flux state in Tab.~\ref{t-fit}). Other model parameters (like the spin, the inclination angle of the disk, and the iron abundance) cannot change value over a few days, which is the timescale of the observations. Fig.~\ref{f-r2} shows the model and the ratio between the data and the best-fit model for every flux state. The constraints on the spin parameter $a_*$ and the Johannsen deformation parameters $\epsilon_3$ are shown in Fig.~\ref{f-c2}.

\begin{table*}
\centering
\begin{tabular}{lccccc}
& GS~1354--645 & Ark~564 & 1H0419--577 & PKS 0558--504 & MCG--6--30--15 \\
\hline
{\sc tbabs} &&&&& \\
$N_{\rm H}$ & $0.7^*$ & $0.0674^*$ & $0.0134^*$ & $0.039^*$ & $0.039^*$ \\
\hline
{\sc relxill\_nk} &&&&& \\
$q_{\rm in}$ & $9.6_{-2.0}$ & $> 9.8$ & $7.6_{-1.1}$ & $> 9.4$ & $6.7_{-2.0}^{+1.6}/7.5_{-1.2}^{+1.3}/7.7_{-0.5}^{+0.5}/8.4_{-0.5}^{+0.6}$ \\
$q_{\rm out}$ & $< 1.4$ & $3^*$ & $= q_{\rm in}$ & $3^*$ & $3^*$ \\
$R_{\rm br}$~[$M$] & $6.9_{-1.6}^{+3.4}$ & $3.093_{-0.117}^{+0.021}$ & -- & $2.81_{-0.20}^{+0.10}$ & $2.8_{-0.8}^{+0.8}/2.89_{-0.59}^{+0.23}/3.26_{-0.42}^{+0.23}/3.3_{-0.4}^{+0.4}$ \\
$a_*$ & $0.959_{-0.013}^{+0.020}$ & $0.997_{-0.008}$ & $0.995_{-0.023}$ & $0.995_{-0.013}$ & $0.964_{-0.012}^{+0.008}$ \\
$\epsilon_3$ & $-1.6_{-1.0}^{+2.1}$ & $0.41_{-0.69}^{+0.11}$ & $-0.2_{-2.1}^{+0.5}$ & $0.0_{-0.8}^{+0.1}$ & $-0.05_{-0.17}^{+0.29}$ \\
$i$~[deg] & $71.9_{-1.1}^{+4.0}$ & $32.2_{-2.0}^{+1.8}$ & $72_{-4}^{+6}$ & $44.4_{-2.9}^{+2.3}$ & $31.6_{-1.6}^{+1.4}$ \\
$z$ & $0^*$ & $0.0247^*$ & $0.104^*$ & $0.1372^*$ & $0.007749^*$ \\
$\log\xi$ & $2.13_{-0.11}^{+0.20}$ & $3.328_{-0.039}^{+0.010}$ & $0.69_{-0.21}^{+0.11}$ & $2.98_{-0.08}^{+0.04}$ & $2.89_{-0.06}^{+0.05}/3.009_{-0.055}^{+0.010}/3.065_{-0.016}^{+0.014}/3.15_{-0.03}^{+0.03}$ \\
$A_{\rm Fe}$ & $0.63_{-0.10}^{+0.06}$ & $4.8_{-0.4}^{+0.7}$ & $2.1_{-0.6}^{+0.4}$ & $5.0_{-0.9}^{+1.9}$ & $3.05_{-0.30}^{+0.24}$ \\
$\Gamma$ & $1.654_{-0.019}^{+0.065}$ & $2.587_{-0.008}^{+0.009}$ & $2.153_{-0.020}^{+0.036}$ & $2.32_{-0.11}^{+0.10}$ & $1.952_{-0.008}^{+0.010}/1.970_{-0.007}^{+,0.008}/2.010_{-0.007}^{+,0.007}/2.021_{-0.007}^{+,0.007}$ \\
$E_{\rm cut}$ [keV] & $144_{-11}^{+19}$ & $300^*$ & $300^*$ & $300^*$ & $196_{-28}^{+28}/155_{-17}^{+23}/163_{-21}^{+28}/269_{-47}^{+68}$ \\
$R_{\rm ref}$ & $0.30_{-0.03}^{+0.05}$ & $3.7_{-0.7}^{+0.6}$ & $1.19_{-0.19}^{+0.32}$ & $1.08_{-0.08}^{+0.21}$ & $0.40_{-0.03}^{+0.04}/0.37_{-0.03}^{+0.03}/0.53_{-0.03}^{+0.05}/0.47_{-0.09}^{+0.03}$ \\
\hline
{\sc xillver} &&&&& \\
$\log\xi'$ &&& $0^*$ && $0^*$ \\
\hline
{\sc zgauss} &&&&& \\
$E_{\rm line}$ [keV] &&&& $6.95_{-0.07}^{+0.08}$ & $0.8142_{-0.0006}^{+0.0007}$ \\
\hline
{\sc zgauss} &&&&& \\
$E'_{\rm line}$ [keV] &&&&& $1.225_{-0.009}^{+0.012}$ \\
\hline
{\sc warmabs$_1$} &&&&& \\
$N_{\rm H \, 1}$ &&&&& $0.56_{-0.06}^{+6.42}/1.166_{-0.049}^{+0.020}/0.994_{-0.030}^{+0.024}/0.25_{-0.06}^{+0.05}$ \\
$\log\xi_1$ &&&&& $1.89_{-0.03}^{+0.04}/1.955_{-0.019}^{+0.014}/1.921_{-0.027}^{+0.018}/2.48_{-0.16}^{+0.20}$ \\
\hline
{\sc warmabs$_2$} &&&&& \\
$N_{\rm H \, 2}$ &&&&& $0.54_{-0.06}^{+6.25}/0.022^*/0.53_{-0.16}^{+0.19}/0.73_{-0.04}^{+0.13}$ \\
$\log\xi_2$ &&&&& $1.88_{-0.06}^{+0.04}/3.1_{-0.6}/3.23_{-0.09}^{+0.06}/1.829_{-0.023}^{+0.021}$ \\
\hline
{\sc dustyabs} &&&&& \\
$\log N_{\rm Fe}$ &&&&& $1.7410_{-0.0030}^{+0.0008}$ \\
\hline
$\chi^2/\nu$ & 2888.41/2722 & 1570.00/1449 & 2488.81/2344 & 1381.05/1311 & 3028.01/2685 \\
& =1.06114 & =1.08351 & =1.06178 & =1.05343 & =1.12775
\end{tabular}
 \caption{Best-fit values of the model parameters of the sources analyzed in this work. $N_{\rm H}$, $N_{\rm H \, 1}$, $N_{\rm H \, 2}$, and $N_{\rm Fe}$ in units $10^{22}$~cm$^{-2}$. $\xi$, $\xi'$, $\xi_1$, and $\xi_2$ in units erg~cm~s$^{-1}$. The reported uncertainties correspond to the 90\% confidence level for one relevant parameter. $^*$ indicates that the parameter is frozen. For MCG--6--30--15, some parameters are supposed to be constant over different flux states and we thus report a single measurement, while other parameters are expected to vary over different flux states and we thus report four measurements (in the order low, medium, high, and very-high flux state).}
\label{t-fit}
\end{table*}

\section{Discussion and conclusions \label{s-dis}}

\begin{figure}[t]
\begin{center}
\includegraphics[width=8.5cm,trim={1.0cm 2.0cm 0cm 1.0cm},clip]{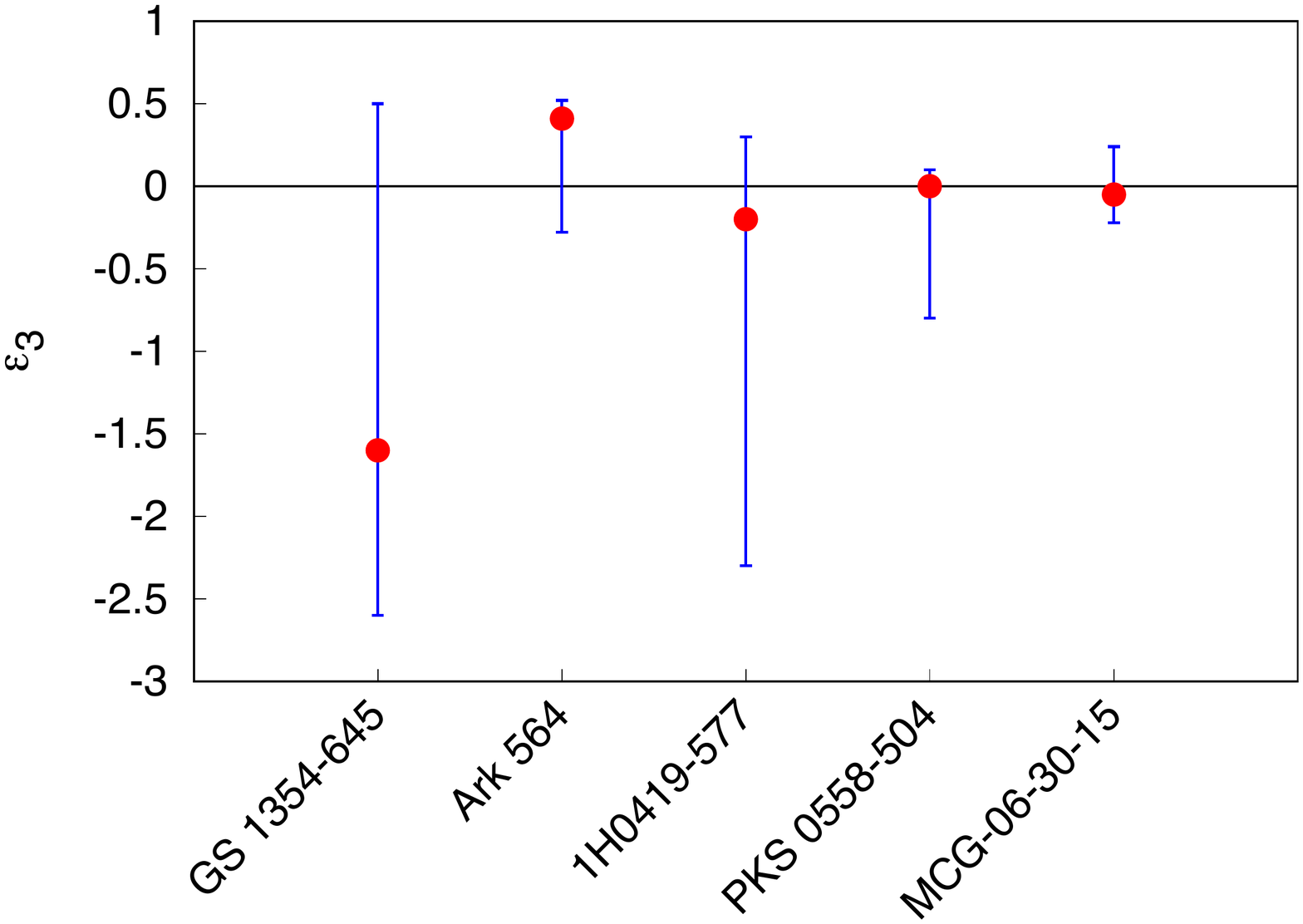}
\end{center}
\vspace{-0.5cm}
\caption{Summary of the measurements of $\epsilon_3$ from the sources analyzed in this work. The reported uncertainties correspond to the 90\% confidence intervals for one relevant parameter. \label{f-all}}
\end{figure}

In this work, we have selected five sources (a stellar-mass black hole and four supermassive black holes) from our previous tests of the Kerr metric using X-ray reflection spectroscopy and we have re-analyzed their data to constrain the Johannsen deformation parameter $\epsilon_3$. Constraints on the spin parameter $a_*$ and the deformation parameters $\epsilon_3$ of the five sources are shown in Figs.~\ref{f-c1} and \ref{f-c2} and all the measurements of $\epsilon_3$ together are shown Fig.~\ref{f-all}.

All our results are nicely consistent with a vanishing value of the deformation parameter $\epsilon_3$; that is, the spacetime metric around these five sources is compatible with the Kerr metric of general relativity already at 90\% confidence level. While such a conclusion was also the result found in our previous work on the measurements of the deformation parameters $\alpha_{13}$ and $\alpha_{22}$, it was not obvious that we will have confirmed that result with the deformation parameter $\epsilon_3$. In general, every non-Kerr feature has its own impact on the relativistic effects of the background metric and on the electromagnetic spectrum of the source, as well as its own correlations with the other parameters of the model.

The five sources analyzed here (GS~1354--645, Ark~564, 1H0419--577, PKS~0558--504, and MCG--6--30--15) had all provided very stringent constraints on $\alpha_{13}$ and $\alpha_{22}$ in our previous work. Here we note some important differences on the strength of the constraint on $\epsilon_3$ from different sources.

$\epsilon_3$ is not well constrained from the data of GS~1354--645 and 1H0419--577. It is remarkable that these are the two sources in which the fit does not suggest a lamppost geometry for the corona. The importance of the emissivity profile on the measurement of the deformation parameter of the metric was already emphasized in~\cite{noi4}. Here, with the deformation parameter $\epsilon_3$ that has a weaker impact on the reflection spectrum than $\alpha_{13}$ and $\alpha_{22}$, it seems that the role of the emissivity profile is even more important and stronger constraints can be obtained when the data require a broken power law with $q_{\rm out} = 3$. Such a conclusion, if correct, should be checked with a larger number of sources, but it is beyond the explorative scope of this work. For GS~1354--645, it is possible that the corona geometry is more complicated and even a broken power law with $q_{\rm out}$ free is not able to discriminate the impact of the intensity profile from the relativistic effects of the spacetime metric.

For the three bare active galactic nuclei (Ark~564, 1H0419--577, and PKS~0558--504), the fits prefer extremely high values of the spin parameter $a_*$. While it is possible that the spin parameters of these black holes have been spun up by prolonged disk accretion, which could lead to approach the famous Thorne bound $a_*^{Th} = 0.998$~\cite{thorne}, the three sources are likely accreting at super-Eddington rate~\cite{md1,md2,md3}. In such a case, their accretion disk is presumably fat and the inner edge of the accretion disk may be inside the ISCO, while in {\sc relxill\_nk} we assume an infinitesimally thin disk with the inner edge at the ISCO radius. For MCG--6--30--15, the mass accretion rate has been estimated to be $0.40 \pm 0.13$ in Eddington unit~\cite{md2}, which can be somewhat higher for the conditions required by a Novikov-Thorne accretion disk, but deviations from a thin disk may be moderate. It is remarkable that MCG--6--30--15 provides the best constraint on $\epsilon_3$ without a measurement stuck at the boundary of the parameter space. We may thus argue that the measurement of $\epsilon_3$ MCG--6--30--15 may be the most reliable among the five here and nicely provides the best constraints.

{\sc relxill\_nk} has a number of approximations, which can be roughly grouped into three classes: $i)$ approximations in the calculations of the reflection spectrum (atomic physics), $ii)$ approximations in the description of the accretion disk (astrophysics), and $iii)$ some relativistic effects are neglected (gravitational physics). In the first class, we can list, for example, the assumption that the electron density of the disk is fixed to $10^{15}$~cm$^{-3}$, the disk is assumed to have Solar elemental abundances with the exception of iron, the calculations assume a cold disk of constant density over height and radius. Among the simplifications in the structure of the disk, we assume that the disk is infinitesimally thin, the inner edge is exactly at the ISCO radius, there is no emission inside the ISCO, the disk is described by a single ionization parameter $\xi$, etc. Concerning the simplification in the gravitational physics sector, most of them have likely a very weak impact on the observed spectrum (like the radiation emitted from the other side of the disk), while other may be important but related to the assumptions of the geometry of the corona. In order to improve our tests of the Kerr metric, we surely need to improve the theoretical model in {\sc relxill\_nk}, but presumably we should also select the right sources/observations, namely those situations that can be better described by our improved theoretical model.

Lastly, we may wonder whether gravitational wave tests can measure $\epsilon_3$ and whether they may provide stronger or weaker bounds with respect to those obtained in our work. As a matter of fact, gravitational waves can be used to test a gravity theory, because the signal is determined by the corresponding field equations, while here we do not have any theory but only a phenomenological black hole metric. If we assume that the quasi-normal mode spectra of scalar and gravitational perturbations are not too different, we can proceed as in Ref.~\cite{roman} and constrain the deformation parameter $\epsilon_3$ from the detection of the frequencies of black hole ringing. In such a framework, we could already obtain constraints on $\epsilon_3$ from the gravitational wave signals observed by LIGO. Considering that the uncertainties on the spin of the final black hole in the events observed by LIGO so far are much larger than the uncertainties on the spins reported in this paper (mainly because here we have selected sources with very high spin), we may expect that the observed gravitational wave events would provide constraints on $\epsilon_3$ weaker than those in this work. Much stronger constraints should instead be expected from the observation of gravitational waves from extreme-mass ratio inspirals~\cite{shuo}, hopefully possible with space-based gravitational wave antennas.


{\bf Acknowledgments --}
This work was supported by the National Natural Science Foundation of China (NSFC), Grant No.~U1531117, and Fudan University, Grant No.~IDH1512060. A.T. also acknowledges support from the China Scholarship Council (CSC), Grant No.~2016GXZR89. A.B.A. also acknowledges the support from the Shanghai Government Scholarship (SGS). S.N. acknowledges support from the Excellence Initiative at Eberhard-Karls Universit\"at T\"ubingen.


\end{document}